# Noise effect on Grover algorithm

Pedro J. Salas[a]

Depto. Tecnologías Especiales Aplicadas a la Telecomunicación, E.T.S.I. Telecomunicación, U.P.M., Ciudad Universitaria s/n, 28040 Madrid (Spain)

ABSTRACT

The decoherence effect on Grover algorithm has been studied numerically through a noise modelled by a depolarizing channel. Two types of error are introduced characterizing the qubit time evolution and gate application, so the noise is directly related to the quantum network construction. The numerical simulation concludes an exponential damping law for the successive probability of the maxima as time increases. We have obtained an allowed-error law for the algorithm: the error threshold for the allowed noise behaves as $\varepsilon_{th}(N) \sim 1/N^{1.1}$ (N being the size of the data set). As the power of N is almost one, we consider the Grover algorithm as robust to a certain extent against decoherence. This law also provides an absolute threshold: if the free evolution error is greater than 0.043, Grover algorithm does not work for any number of qubits affected by the present error model. The improvement in the probability of success, in the case of two qubits has been illustrated by using a fault-tolerant encoding of the initial state by means of the [[7,1,3]] quantum code.

## 1. INTRODUCTION

The discovery of Shor's algorithm opened up Pandora's Box of quantum computation. Other algorithms followed it, although they did not change the classic complexity class, the quantum version runs faster than any known classic algorithm. This is the case of quantum search algorithm proposed by Grover [1]. Nevertheless when considering the physical implementation of these algorithms, it was discovered that decoherence and dissipation (as spontaneous emissions) were going to be the true bottlenecks that would limit the usefulness of the quantum algorithms.

The Grover algorithm has been successfully implemented, see for example [2]. Nevertheless, the algorithm requires the synthesis and handling of highly entangled states that are very prone to decoherence. Some attempts at simulating the effect of errors in the algorithm have been published. Pablo-Norman and Ruiz-Altaba [3] introduce gaussian white noise into each step of the algorithm considering a description of the algorithm as a rotation in a two-dimensional space. The model only takes into account two different types of error, one affecting the required state and other for the orthogonal state. The noise does not modify the number of iterations at which the maxima appear ($k_{max} \sim \lfloor N^{1/2} \pi/4 \rfloor$) with respect the case of no noise, although their probabilities are now smaller. They conclude that the allowed noise law for the algorithm scales as $N^{-2/3}$ ($N = 2^n$ being the size of the data list). Using a two-state model and representing the time evolution of the algorithm through an SU(2) transformation,

---

[a] Electronic mail psalas@etsit.upm.es

Yu and Sun [4] have studied the effect of the decoherence induced by the environment interaction, concluding the slight robustness of the algorithm against the noise.

In the first stages of carrying out the algorithm, the gate imperfections dominate the evolution. At longer times, the decoherence determines the results. Long *et al.* [5] study the effect of the gate imperfections on the Grover algorithm without decoherence. They conclude that the size of the database is limited by a law of the power minus two of a parameter that measures the gate imperfection. Hsieh *et al.* [6] indicate that Long underestimates the allowed error in a factor $2^{1/2}$, although they maintain the dependency with N. Shapira *et al.* [7] study the effect of the unitary noise characterized by standard deviation ε that must fulfil the condition $\varepsilon < O(n^{-1/2}N^{-1/4})$ to maintain a significant efficiency.

Chen *et al.* [8] modelled the noise in the algorithm by means of the depolarizing channel. Only evolution errors are taken into account through the density matrix formalism, achieving an equation that allows the evolution of the probability of the required state to be represented depending on the time step considered. On the other hand, Song and Kim [9] carried out a similar study by means of two models: a first stochastic model with two levels and dissipation and a second model with unitary imperfections. Both models agree after a suitable adjustment of their parameters. The results of the second model can be understood through two mechanisms: a stochastic rotation within a two-dimension subspace ($\mathcal{H}_2$) of the total Hilbert space in which is developed the algorithm and another mechanism of diffusion towards the complete Hilbert space ($\mathcal{H}_N$). The probability of the system remaining within $\mathcal{H}_2$ decreases exponentially over time, characterized by a decoherence parameter γ. The average fidelity value decreases over time approximately in an exponential form.

Ellinas and Konstadakis [10] consider the effect of the decoherence by interaction with the environment of the algorithm with two states, concluding its robustness, being able to make searches successfully after $N^{1/2}$ applications of Grover gate. Azuma [11] introduces phase errors in each qubit and time step. Making a perturbative development, he calculates the terms numerically until fifth order, and explores the region in which the algorithm with noise finds the required item with a probability threshold ($p_{th}$) after applying M gates. The conclusion is that the allowed noise behaves as $1/M(1-p_{th})n$.

The effect of a noisy oracle in the algorithm is studied by Shenvi *et al.* [12]. They use a discrete and continuous model to introduce random phase errors. They find that if the size of the oracle increases according to a factor k, the error must decrease as $k^{1/4}$ to maintain the probability of success.

The chaotic behaviour coming from the static interactions between qubits has been studied by Pomeransky *et al.* [13] and the dissipation coming from non-unitary errors, by Zhirov *et al.* [14].

In the previous studies the noise is not introduced explicitly into the quantum gates, since they neither consider an explicit implementation nor the growth of the total error with the complexity of the circuit, which depends on the number of gates as well as the parallelism that implements the Grover gate.

In this work, we will study the robustness of Grover algorithm against the noise. The noisy algorithm will be numerically simulated by means of the isotropic depolarizing channel model. The noise is introduced by means of two parameters (ε, γ) related to the free evolution and gate error probabilities. Section 2 summarizes the main steps of Grover algorithm and the quantum networks that implement it. Section 3 establishes the assumption of the decoherence model. Section 4 puts together the numerical results concluding several effects of the noise: an exponential-time damping law, a displacement for the successive maxima and an allowed-error law. Finally the quantum error correcting codes usefulness is shown by means of a simple binary [[7,1,3]] fault-tolerant encoding.

## 2. GROVER SEARCH ALGORITHM

Any classic algorithm for searching an item in a randomly ordered database of N entries requires O(N) steps on average. Grover discovered a quantum algorithm [1] that runs in $O(N^{1/2})$ steps. Let us review the protocol.

Suppose we wish to search through an unstructured database with N items. Rather than search for the items directly, we concentrate on the *index* of those items, which is just a number within the range 0 to N-1. For convenience we assume $N=2^n$, so the index can be stored in n bits. By definition, the solution can be represented by means of a function f such as: $f(x_s) = 1$ if $x_s$ is a solution to the search problem and $f(x) = 0$ if x is not a solution. A classic algorithm would need to calculate (N-1) values of the function f(x) to obtain the solution with certainty (assuming that $x = x_s$ exists), and the number of computational steps increases as O(N).

Using quantum mechanics, Grover showed that it is possible to decrease the number of f-calls. Suppose the unstructured database with N items DB = {$x_0$, …,$x_{N-1}$}, and we are searching for the item $x_s$ so $f(x_s) = 1$ and $f(x_k) = 0$ $\forall k \neq s$. The quantum Grover algorithm has the following steps:

1) *Synthesis of the state superposition* of all indices $|\Psi_0\rangle$. Apply the n-qubit Hadamard transformation to the initial state $|0^{\otimes n}\rangle$

$$H^{\otimes n} = H \otimes ..\overset{(n)}{..} \otimes H \quad \text{with} \quad H = \frac{1}{\sqrt{2}}\begin{pmatrix} 1 & 1 \\ 1 & -1 \end{pmatrix} \quad (1)$$

$$|\Psi_0\rangle = H^{\otimes n}|0^{\otimes n}\rangle = \frac{1}{\sqrt{N}}\sum_{x=0}^{N-1}|x\rangle = \text{index state superposition} \quad (2)$$

2) Application of *Grover gate* $G = (2 H^{\otimes n} |0^{\otimes n}\rangle\langle 0^{\otimes n}| H^{\otimes n} - I_n) (I_n - 2 |x_s\rangle\langle x_s|) = -I_{|\Psi_0\rangle} I_{|x_s\rangle}$, where $I_n$ is the identity of dimension n and $I_{|\phi\rangle} = I_n - 2|\phi\rangle\langle\phi|$.

The first part of G is the inversion with respect to the average of the coefficients and the second part inverts the sign of the required item $x_s$ and is functioning as an efficient "black box" called oracle:

$$I_{|x_s\rangle}|x\rangle = (-1)^{f(x)}|x\rangle \begin{cases} -|x_s\rangle & \text{if } x = x_s \\ |x\rangle & \text{otherwise} \end{cases} \quad (3)$$

Rewriting the $|\Psi_0\rangle$ state in terms of the orthonormal basis {$|x_s\rangle$, $|x_s^\perp\rangle = \sum_{x \neq x_s}|x\rangle/\sqrt{N-1}$}:

$$|\Psi_0\rangle = cos(\theta/2)|x_s^\perp\rangle + sin(\theta/2)|x_s\rangle \quad \text{with} \quad sin(\theta/2) = 1/\sqrt{N} \quad (4)$$

and the successive application of Grover gate provides:

$$G^k |\Psi_0\rangle = \cos([2k+1](\theta/2)) |x_s^\perp\rangle + \sin([2k+1](\theta/2)) |x_s\rangle \tag{5}$$

For huge values of N, $\sin(\theta/2) \sim \theta/2$ and the probability of success will be $P_s = |\sin([2k+1](\theta/2))|^2 \sim 1$ when $k = k_0 \sim \lfloor \pi N^{1/2}/4 \rfloor \sim O(N^{1/2})$.

3) After $k_0$ applications of the Grover gate, we measure on the state $G^{k_0} |\Psi_0\rangle$

The conclusion is that after the $O(N^{1/2})$ Grover gate calls, the measurement on the state provides a probability near to one to obtain the required item $x_s$.

## 2.1 Quantum networks implementing the algorithm

In order to introduce the errors into the algorithm we implement it as a quantum network Q. The following over-complete gate set is used { H (Hadamard), X and Z (Pauli gates), CNOT = $C^2(X)$, T(Toffoli) = $C^3(X)$ } according the simplicity criterion of not using gates involving more than three qubits. The CNOT gate only is used in the case of n=2, in the remaining cases only one and three qubit gates are involved. In spite of the network implementing the Grover gate G has two separate pieces, the oracle $I_{|x_s\rangle}$ = ($I_n - 2 |x_s\rangle\langle x_s|$) and the inversion $-I_{|\psi_0\rangle}$ = ($2 H^{\otimes n} |0^{\otimes n}\rangle\langle 0^{\otimes n}| H^{\otimes n} - I_n$), their construction is closely related. Evidently, the gate set chosen is not the minimum but the main goal is to achieve scaling laws for the permitted error that should be, to a certain extent, independent of the set used. Note that all the results obtained will be strictly applicable to the present gate and error model.

The initial state $|\Psi_0\rangle$ is synthesized rotating Hadamard the n-qubit initial register $|0^{\otimes n}\rangle$. When no error is present in the algorithm, the result of the algorithm does not depend on the state $|x_s\rangle$ chosen but in the noisy case it does through the oracle piece of the network. Choosing the noisiest $|x_s\rangle$ state will provide an upper bound of the error network. The noisiest $|x_s\rangle$ will be the one whose network involves more time steps and gates, so we chose $|x_s\rangle = |0^{\otimes n}\rangle$. The general pieces of network implementing the oracle and the inversion $-I_{|\psi_0\rangle}$ are shown in figure 1. They are constructed by means of generalized control-Z gates involving (n-1) control qubits and the n-th as the target qubit. These gates can be transformed into generalized Toffoli gates as $C^n(X) = H(n) C^n(Z) H(n)$, H(n) being a Hadamard gate applied on the n-th qubit. Figure 2 shows the breakdown of these $C^n(X)$ gates into the $C^3(X)$ gates considered in the universal set. Note the $C^n(X)$ implementation involves some additional ancilla qubits.

Taking the above networks into account it is not difficult to show how fast the resources increase. The number of Toffoli gates are 2(n-2) ((n-2) for the oracle and (n-2) for the $-I_{|\psi_0\rangle}$ piece of network, n>2); 3n Hadamard gates, 2(2n-1) X gates and 2 Z gates; these make 7n one-qubit gates. The number of total gates increases as O(n). The ancilla qubits necessary to implement the $C^n(X)$ gates are (n-3), i.e. O(n). The number of total time steps of the algorithm is 2n+6, increasing as O(n).

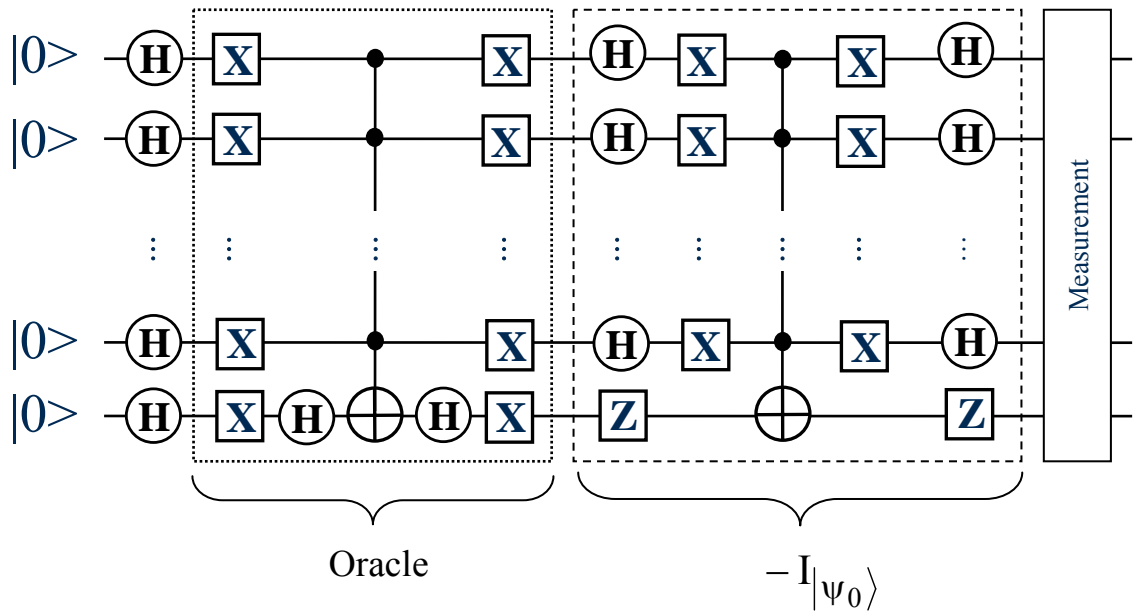

Figure 1. General network implementing Grover algorithm when the oracle is searching for the noisiest state $|0^{\otimes n}\rangle$.

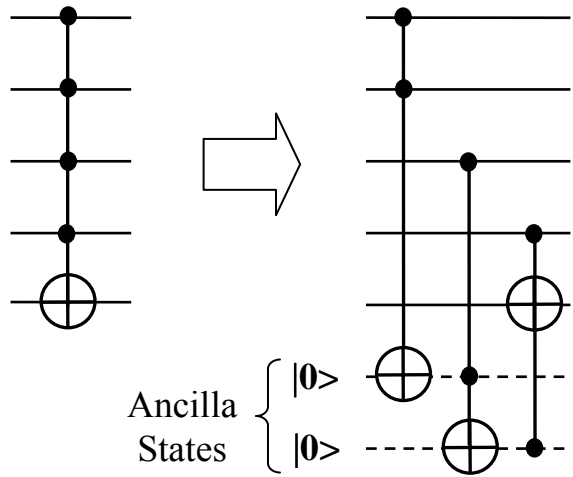

Figure 2. Implementation of a general $C^n(X)$ gate by means of Toffoli gates.

## 3. DECOHERENCE MODEL

To simulate the noisy quantum network Q an *independent stochastic error model* [15] based on the notion of error locations [16] is used. In a given location or gate of the network a random error is introduced. Each error is independent of the other errors happening at the same or different locations. All quantum steps have some probability of error, and we distinguish between *memory errors* (resulting from qubit free-evolution) with error probability ε and *one-qubit* and *two-qubit gate errors* with probability of error *proportional* to a parameter γ.

Memory errors are located at each time step in the network, affecting all the qubits evolved in that step. Their effect can be controlled constructing highly parallelized networks. To model the evolution errors we consider the depolarizing channel model. For each error location affecting one qubit with an ε probability of error, we consider an isotropic ε/3 error probability for the X, Y and Z, as long as the probability of having no evolution error is (1-ε).

We consider two ways in which the noise affects the gates: we assume each gate is implemented in one time step, so an evolution error with ε error probability is introduced into all the qubits in addition to an intrinsic gate error with a probability proportional to γ affecting the qubits involved in the gate application. For the noisy one-qubit gates (Hadamard, Pauli and measurement), γ is the error probability set up at the gate location (with γ/3 isotropic for X, Y and Z). In the two-qubit gate (CNOT, in the n=2 case), we assume there are sixteen possibilities corresponding to the tensor product {I, X, Y, Z} ⊗ {I, X, Y, Z}. If the one qubit gate error probability is γ, each two-qubit error appears with probability γ/15, because the I⊗I term is not, actually, an error operation. We let the gate operate before the error proceeds. This O(γ) (instead of O($\gamma^2$)) two-qubit error behaviour accounts for the possible gate error spreading. Note that in this model γ only accounts for the error coming strictly from the gate application. For the three-qubit Toffoli gate, we assume there are sixty four possibilities corresponding to the tensor product {I, X, Y, Z}$^{\otimes 3}$ and the error is γ/63 (instead O($\gamma^3$)) because $I^{\otimes 3}$ is not an error. As with CNOT, we assume the Toffoli gate is carried out in one time step affected by an error probability of ε per qubit and the gate operates before the error proceeds. Note that, even though the γ values are the same for one, two and three qubit gates, the effective error is different for each gate.

Neither leakage errors nor explicit assumptions on scaling problems are taken into account. We assume that ε and γ errors are independent of the total number of network qubits.

Errors are introduced into the calculation using the Luxury Pseudorandom Numbers [17] which is an improvement of the subtract-and-borrow random number generator proposed by Marsaglia and Zaman. The fortran-77 code is due to James [18], and is used with the luxury level parameter p = 223. As the code state for this value of p, any theoretically possible correlations have very small chance of being observed. The code returns a number of 32-bit random floating point number in the range (0, 1). For each run a new random seed is chosen as a 32-bit integer.

The noiseless quantum network Q can be represented by means of a quantum operator $\hat{Q}$ which is a sequence of time step operators $\hat{T}_i$ and gates $\hat{G}_j$, $\hat{Q} = \hat{T}_t \circ \hat{G}_t \circ \cdots \hat{T}_1 \circ \hat{G}_1$. Each time step has the structure $\hat{T}_i = \hat{I}^{\otimes n}$ and $\hat{G}_j$ involves a tensor product of some unitary gates ($\hat{g}_j(q_{\{k_j\}})$) affecting a subset of $\{k_j\}$ qubits and the identity operators affecting the remaining qubits: $\hat{G}_j = \hat{g}_j(q_{\{k_j\}}) \otimes_{s \neq \{k_j\}}^n \hat{I}_s$. The noise transforms $\hat{Q}$ into $\hat{Q}_{noisy}$, where each $\hat{T}_i$ has the new form $\hat{T}_{i,noisy} = \hat{A}_{i_1}^1 \otimes \cdots \otimes \hat{A}_{i_n}^n$, and {$\hat{A}_{i_k}^k$, k=1,…,n} ∈ {$\hat{I}(i_k = 0), \hat{X}(i_k = 1), \hat{Y}(i_k = 2), \hat{Z}(i_k = 3)$} each with probabilities (1-ε), ε/3, ε/3 and ε/3 respectively. The new form of $\hat{G}_{j,noisy} = \hat{g}_{j,noisy}(q_{\{k_j\}}) \otimes \hat{T}_{j,noisy}$, where one noisy time step is introduced ($\hat{T}_{j,noisy}$) and the corresponding noisy gates ($\hat{g}_{j,noisy}$) affecting the $q_{\{k_j\}}$ qubits with error probabilities O(γ). Each error distribution among the network provides a different noisy quantum path.

We search for the noisiest state, involving the largest number of gates $\rho_0 = |0^{\otimes n}\rangle\langle 0^{\otimes n}|$, and the final noisy state $\rho_f$ is the weighted average over the output density matrices for each noisy quantum paths ($\hat{Q}_{noisy} \circ \rho_0$), according their probabilities $P_{\hat{Q}_{noisy}}(\varepsilon,\gamma,n)$:

$$\rho_{noisy} = \sum_{\substack{noisy \\ paths}} P_{\hat{Q}_{noisy}}(\varepsilon,\gamma,n)(\hat{Q}_{noisy} \circ \rho_0) \tag{6}$$

The final success probability of the noisy Grover algorithm is:

$$P_S(\varepsilon,\gamma,n) = \langle \text{searched state}|\rho_{noisy}|\text{searched state}\rangle = \langle x_s|\rho_{noisy}|x_s\rangle \tag{7}$$

From the numerical point of view, these probabilities are calculated as:

$$P_S(\varepsilon,\gamma,n) = \frac{1}{N_C}\sum_{i=1}^{N_C}\langle 0^{\otimes n}|\rho_{noisy,i}(\varepsilon,\gamma,n)|0^{\otimes n}\rangle \tag{8}$$

$N_C$ being the number of total calculations. Equation (8) is a statistical approximation to equation (7), the first being exact when $N_C \to +\infty$. In order to reach the numeric convergence of the probability $P_S$, the value of $N_C$ is taken fulfilling the condition $N_C \geq 10 \times \max(1/\varepsilon, 1/\gamma) = N_0$, checking (by comparison of $P_S$ values when $N_C \gg N_0$) that this choice of $N_C$ assures a convergence in the $P_S$ bigger than the 0.5%.

## 4. NUMERICAL RESULTS

With the previous decoherence model we have numerically simulated the noisy Grover algorithm, studying several aspects depending on the values of $\varepsilon$, $\gamma$ and n.

### 4.1 Exponential-time damping law

The most evident effect of the noise on the algorithm is the damping of the maxima for the success probability ($P_S$) as the time increases. We have carried out a variety of calculations for $P_S(\varepsilon,\gamma, n=2,..,7)$ for several $\varepsilon$ and $\gamma$ values, confirming the dependence law:

$$P_S(\varepsilon,\gamma,n;t) = A(\varepsilon,\gamma,n)e^{-\lambda(\varepsilon,\gamma,n)t} + \frac{1}{2^n} \tag{9}$$

in which the time evolution is strictly exponential. The parameter t does not represent the real time steps in the network in figure 1, but the number of Grover gate (G) applications. The term $1/2^n$ originates because the states involved in the algorithm are the linear combination of all the qubit states having length n, so without dissipation the final population does not vanish to zero when t goes to infinity.

The $A(\varepsilon,\gamma,n)$ is a lengthy function and smaller than one for all the cases studied, so the bigger dependence is included in the function $\lambda(\varepsilon,\gamma,n)$ that we successfully fit to:

$$\lambda(\varepsilon,\gamma,n) = (18.63\ \varepsilon + 8.124\ \gamma)\ n - 5.871\ \varepsilon - 12.336\ \gamma \tag{10}$$

fulfilling $\lambda(\varepsilon,\gamma,n) \geq 0$ for $n \geq 2$ and $0 \leq (\varepsilon, \gamma)$. Note that in case of $\varepsilon = \gamma = 0$ there is no noise and $\lambda = 0$. By fixing our attention in the exponential dependence, equation (9) for $P_S$ can be written in terms of the probability for the first maximum ($P_S(\varepsilon,\gamma,n;t_1)$) reached when $t = t_1 \sim \lfloor \pi N^{1/2}/4 \rfloor$:

$$P_S(\varepsilon,\gamma,n;t) - \frac{1}{2^n} = P_S(\varepsilon,\gamma,n;t_1)e^{-\lambda(\varepsilon,\gamma,n)(t-t_1)} \qquad (11)$$

allowing us to interpret $\tau = 1/\lambda$ as a decoherence parameter. In addition to this exponential-time damping-law for the maxima, the minima increases their probability, reaching the steady limit $P_S(\varepsilon,\gamma,n;t) \sim 1/2^n$ for large enough time parameter (t) values. As an example of this behaviour, figure 3 shows the case of n = 4 when $\varepsilon^{-1} = 3000$ and $\gamma^{-1} = 5000$. The maxima are successfully fitted to the exponential law (11) with $\lambda_{fit} = 0.0282$, while the $\lambda$ value calculated using the equality (10) is 0.0269.

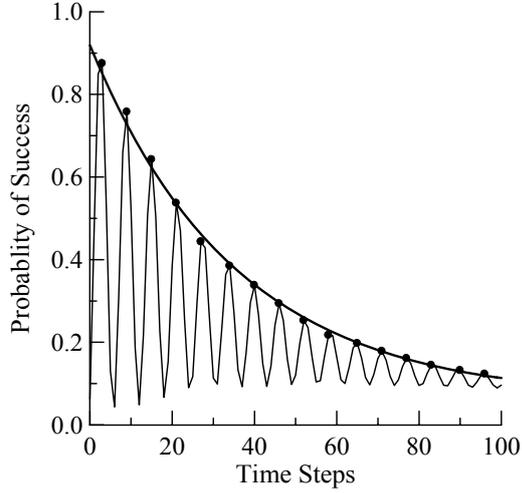

Figure 3. Damping of the probability of success ($P_S$) versus time for noisy Grover algorithm having an $\varepsilon^{-1} = 3000$ evolution error and $\gamma^{-1} = 5000$ gate error, for the case of n=4. Solid line on the maxima corresponds to the fitting curve obtained from equation (10).

4.2 First maximum evolution

If we are interested looking for an item in an unsorted database using Grover algorithm, the minimum number of iterations corresponds to the first maximum and reaching it, we stop the process. In this section we study the noise dependence of the function $P_S(\varepsilon,\gamma,n; t_1)$ for $t_1$ fixed to the first maximum time step. We can reduce the number of error variables without loosing the richness of the results assuming the error condition $\varepsilon = \gamma$. With this condition we have calculated the iteration time at which the first maximum appears. Figure 4 shows the $\varepsilon$-value dependence of the time at which the first maximum appears. Results for the algorithm with n = 4,…,7, qubits show a strong dependence with $\varepsilon$. It is possible to fit the discontinuity points onto a logarithmic curve as shown for n=5, 6 and 7 in figure 4: first maximum time ~ $A(n) \ln \varepsilon + B(n)$.

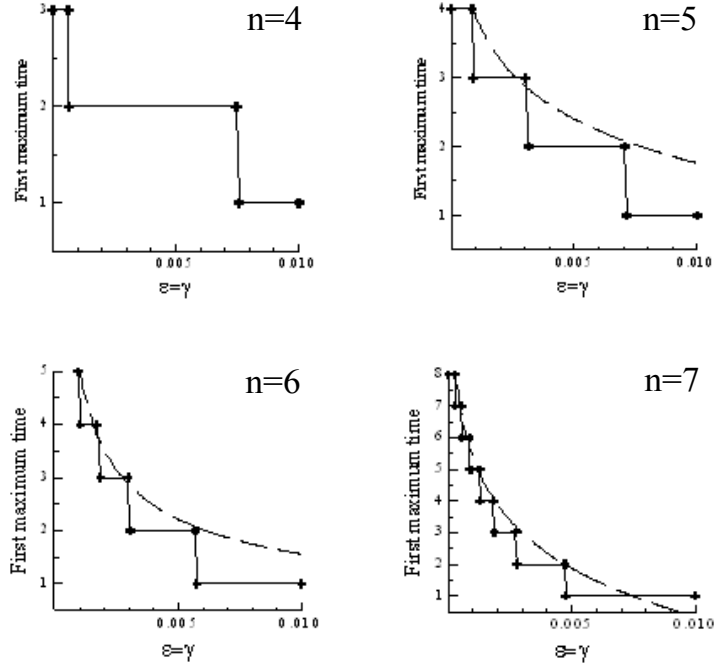

Figure 4. Results for the ε-value dependence of the time at which the first maximum appears for the algorithm with n = 4,…,7 qubits. Solid line for n=5, 6 and 7, fits the discontinuity points to a logarithmic curve $A(n) \ln \varepsilon + B(n)$.

This behaviour shows the surprising fact that the bigger error ($\varepsilon = \gamma$) the smaller number of time steps (Grover gate applications) are required to reach the first maximum, so less computational effort is needed. Its origin is the increasing damping effect on the probability over time, producing a bigger decrease in probability as the time increases. At the same time, the corresponding probability of success for the first maximum decrease will be shown in the next section. The effect can be understood studying the evolution of the coefficients in the state $G^t|\Psi_0\rangle$ (that is the linear combination of $2^n$ n-qubit states) as the time t increases. As an example, a calculation with n = 5 qubits and $\varepsilon^{-1} = \gamma^{-1} = 2000$, has been carried out (see figure 5). For this error, figure 4 shows that the first maximum appears for t = 4. The state $G^t|\Psi_0\rangle$ without noise only involves two different coefficients (see equation 5): $\sin([2k+1]\theta/2)$ for the searched state and $\cos([2k+1]\theta/2)$ (always the same) for the remaining states. Noise destroys this behaviour and the coefficients of the n-qubit states become different. Figure 5 shows the square of the coefficient for each n-qubit state at different time steps. Some local maxima appear for 2, 3, 5, 9 and 17 corresponding to n-qubit states with weight one. Without going into detail on the network implementing the algorithm, the noise effect could be viewed as some errors affecting the noise-free state $G^t|\Psi_0\rangle$ at the corresponding time step t. Concretely, the effect of bit-flips is to interchange the n-qubit states inside $G^t|\Psi_0\rangle$. A bit-flip error of weight w (noted as $X_w$, w having "1" at the positions where the bit-flips occur) appear with a probability $O(\varepsilon^w, \gamma^w)$ and transform the state $G^t|\Psi_0\rangle$ (as well as each n-qubit state) into $X_w G^t|\Psi_0\rangle$. As the more probable errors are for w = 1, the n-qubit states in $X_w G^t|\Psi_0\rangle$ with weight one (remember that the searched for n-qubit state is $|00…0\rangle$) will show a local maximum in its probability (square of the coefficient). The example studied has n = 5, and the maxima for the time steps t = 4 (first maximum), 6 and 8 (first minimum) are shown in figure 5. This maxima pattern is maintained as the time increases. The n-qubit states with weight two or more, have decreasing coefficients depending on the weight of the n-qubit state. If the time increases even more, all the squared coefficients converge to $1/2^n$ and the noise completely destroys the advantage of the algorithm. In fact, if the error is big enough, the probability for the n-qubit searched state could not be the biggest one.

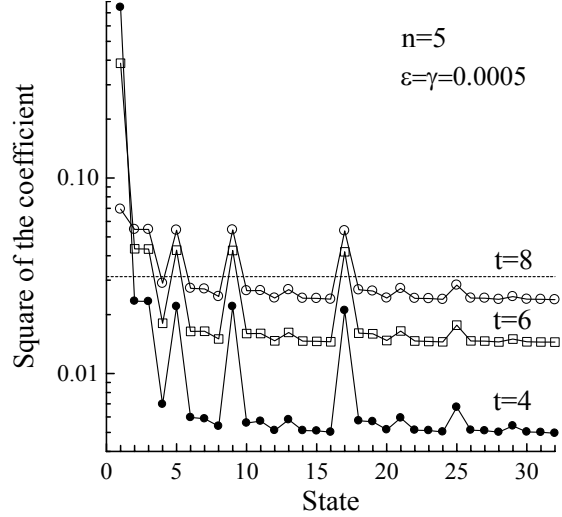

Figure 5. Square of the coefficients for the Grover state with n=5 qubits and $\varepsilon^{-1} = \gamma^{-1} = 2000$ vs. the 5-qubit state number. Some pattern of maxima appears that is maintained when the time increases from t = 4 ● (first maximum), 6 □ and 8 ○ (first minimum). Dashed horizontal line represents the value 1/32, that is the convergent value in the limit of a long enough time evolution.

The effect of the noise in the Grover state is to produce a probability flux, mainly, from the searched for n-qubit state to the remaining ones, this flux being as big as the error increases. In the present work, the searched for n-qubit state is the |00…0⟩ and the noisy $G^t|\Psi_0\rangle$ has local maxima on the n-qubit states with weight one. If the searched for n-qubit state would have weight u, the local maxima in the noisy $G^t|\Psi_0\rangle$ state would appear on n-qubit states having weights u ± 1. This increasing damping of the probability for the searched for n-qubit state, make decrease the time step at which the maxima appear.

### 4.3 Allowed-error law

The most interesting point to be answered is what is the law that limits the allowed error in the algorithm. As was mentioned in the introduction, several results seem to point a $N^{-a}$ law. If a is less or close to 1, we consider the algorithm as robust, other values for a (appreciably greater than 1) will make the algorithm non-robust.

To look for the allowed-error law we study the probability of success ($P_S$) for the first maximum depending on the ε and γ values through its relationship $C = \varepsilon/\gamma$. For each number of qubits (n = 2,…,7) and C, we calculate the probability $P_S(\varepsilon,C,n; t_1)$ and obtain the threshold value $\varepsilon = \varepsilon_{th}(\varepsilon,C,n)$ solving the equation $P_S(\varepsilon,C,n; t_1) = P_{th}$. The value of $P_{th}$ (threshold probability) is chosen $P_{th} = 0.5$ (although it could be any other value). If the condition $\varepsilon < \varepsilon_{th}(C,n)$ is fulfilled, then the $P_S(\varepsilon,C,n; t_1) > P_{th} = 0.5$. As an example of the kind of curves obtained, we present the case C = 1 in figure 6. The crossing points between the curves $P_S(\varepsilon,C=1,n; t_1)$ with the horizontal line $P_{th} = 0.5$, provide the thresholds $\varepsilon_{th}(C=1,n)$. In the case of n = 3, a change of slope is appreciated at the point indicated with an arrow. It originates because at this $\varepsilon \sim 3.4 \times 10^{-3}$ the first maximum change the time step at which it appears from t=2 to t=1, as was mentioned in section 4.2.

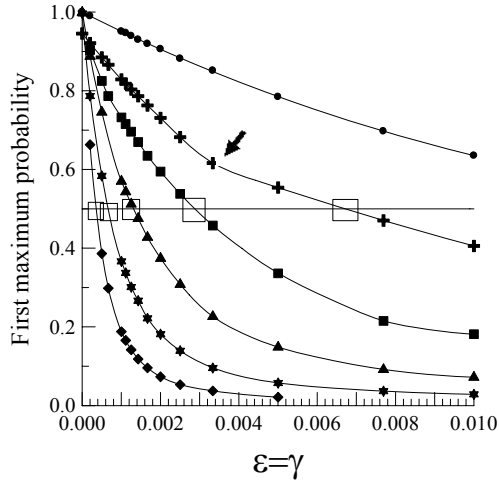

Figure 6. Probability for the first maximum versus $\varepsilon$ ($P_S(\varepsilon,C=1,n; t_1)$) for different n values: 2 ●, 3 ✚, 4 ■, 5 ▲, 6 ★, 7 ♦. Boxes reflect the crossings points between the curves $P_S(\varepsilon,C=1,n; t_1)$ with the horizontal line $P_{th} = 0.5$ and provide the thresholds $\varepsilon_{th}(C=1,n)$. A change of slope is appreciated in the point indicated by an arrow.

Representing the $\ln(\varepsilon_{th}(C,N=2^n))$ versus $\ln N$ we obtain a linear behaviour:

$$\ln(\varepsilon_{th}(C,N=2^n)) = -a(C) \ln N - b(C) \quad (12)$$

The lines for different values of C are parallel, so coefficient a (almost) does not depend on C as shown in figure 7. The a(C) and b(C) are plotted in figure 8, providing an almost constant a(C) value and upper bounded by $a(C=\infty) \sim 1.1$, while b(C) is a slowly varying function.

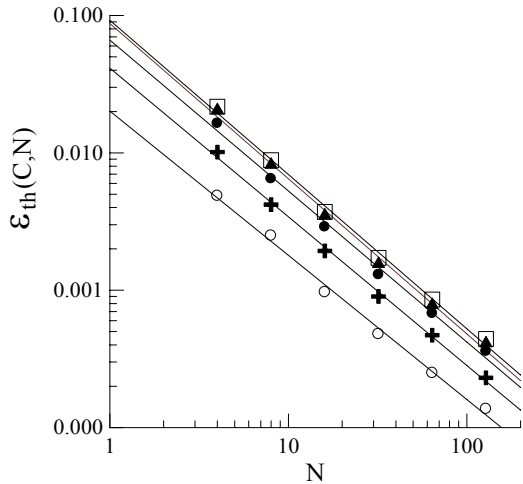

Figure 7. Logarithm of the error threshold $\varepsilon_{th}(C,N)$ versus $N=2^n$ for several C values: 0.1 ○, 0.3 ✚, 1. ●, 6. ▲, ∞ □.

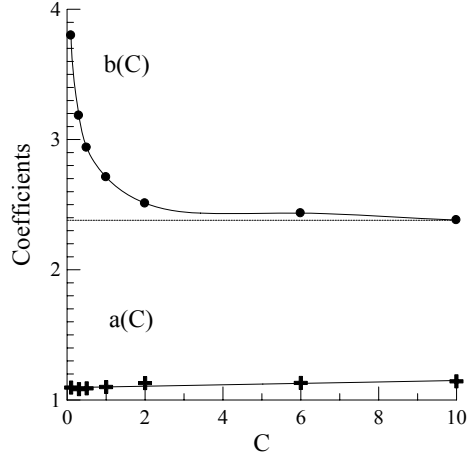

Figure 8. Dependence of the coefficients with C of the law $\ln(\varepsilon_{th}(C,N=2^n)) = -a(C)\ln N - b(C)$.

Equation (12) provides some simple consequences and estimations of the algorithm. Given some hardware characterized by the pair $(\varepsilon, \gamma)$ with $C = \varepsilon/\gamma$, the error threshold for the allowed noise behaves like $\varepsilon_{th}(N) \sim 1/N^{1.1}$ as the size of the data set N increases. Evidently the relationship with the number of qubits exponentially decreases $\varepsilon_{th}(n) \sim e^{-dn}$ if $a = d/\ln 2$ and $d = 0.762$. This law will permit us to establish an absolute threshold for the Grover algorithm if the present error model is fulfilled. If $N = 2^n$, equation (12) provides the maximum number of qubits the algorithm can handle, maintaining the success probability greater than $P_{th} = 0.5$:

$$n \leq \frac{1}{a(C)\ln 2}\left(-b(C) - \ln\varepsilon_{th}(C,N)\right) \qquad (13)$$

In order for $n \geq 1$, the condition $\varepsilon_{th} \leq e^{-b(C)-a\ln 2}$ must be fulfilled. Its maximum value ($\varepsilon_{th}^{max}$) corresponds to the minimum value of $b(C)$, $b(C)_{min} = b(C=\infty) = 2.3802$, and taking $a = 1.1$, we obtain $\varepsilon_{th}^{max} = 0.043$. This value means the greatest error permitted in the present implementation of the algorithm. If $\varepsilon > \varepsilon_{th}^{max}$, the algorithm does not work for any number of qubits affected by the present error model.

We can estimate the maximum number of qubits for a given error $\varepsilon$. Assuming $\varepsilon \sim \gamma = 10^{-5}$ as a possible error, this law provides $\ln(\varepsilon_{th}(C=1,N=2^n)) = -a(1)\ln N - b(1) = -1.1\ln N - 2.711$. For this C=1, the maximum number of qubits is $n \leq 11$ and the size of the data base is $N = 2048$. Unfortunately the size of this database is too small to be of practical interest. In fact, the important searching problems could involve database sizes, for instance, of 56 bits or more, then including $2^{56} \sim 10^{17}$ items in the data set. In this case, assuming C=1, $\varepsilon_{th} \leq e^{-b(C)-na\ln 2} \sim 10^{-20}$, which is completely inaccessible from the experimental point of view. The only way to use the Grover algorithm in real and interesting problems will require a decoherence control method.

### 4.4 Two-qubit encoded Grover algorithm

To control the decoherence a quantum error correcting code could be used. The effect can already be appreciated in the case of two qubits. If we consider the Grover gate application as a compact block, the encoded network is simplified considerably as is pointed out in figure 9. After a fault-tolerant encoding of the |00> state, the $H^{\otimes 2}$ gate is applied, followed by one Grover gate application. After that, an error recovering step

would be required, but as the next step will be a measurement, this quantum correction is not strictly necessarily and could be replaced by a measurement and a classic error correction before identifying the final state. In this context, the code is used more as a passive method to control the decoherence than as an active method, correcting errors.

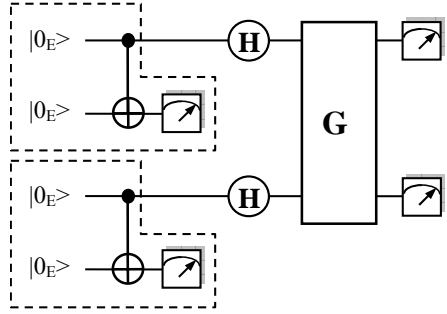

Figure 9. Encoded two-qubit Grover network algorithm. The initial $|0_E\rangle$ states are encoded by means of the [[7,1,3,]] Steane code and are fault-tolerant bit-flip checked using a CNOT gate to another $|0_E\rangle$ state (dashed boxes).

An appropriate encoding must be able to implement the gates involved in the network transversally as shown in figures 9 and 10 for Hadamard and CNOT, so a suitable possibility is the binary Steane CSS code [[7,1,3]].

In order to synthesize the initial encoded $|0_E\rangle \otimes |0_E\rangle$ state fault-tolerantly, we take advantage of some beneficial properties of the error equivalence [19] in the [[7,1,3]] code. It is assumed in standard notation that the classic codeword $u \in GF(2)^7$ produces the quantum state $|u\rangle \in \mathcal{H}^{\otimes 7}$ ($\mathcal{H}$ is the qubit Hilbert space). The quantum code [[7,1,3]] is obtained from a classic Hamming code $C = [7,4,3] \supset C^\perp = [7,3,4]$. The encoded state $|0_E\rangle$ is the linear combination of the states coming from the eight classic $C^\perp$ codewords. The bit-flip errors affecting these states can be stated as $X_v$, v having "1" at the positions where the bit-flips occur. An error is said to have weight $W(X_v) = W_H(v)$, $W_H$ being the usual Hamming weight of $v \in GF(2)^7$. Taking the classic codewords involved in $|0_E\rangle$ as the starting point, the complete Hilbert space $\mathcal{H}^{\otimes 7}$ can be covered considering bit-flip errors of weight $W(X_v) \leq 3$. The Hilbert space $\mathcal{H}^{\otimes 7}$ is partitioned into 16 sets, each having a different four-bit syndrome, considered as codewords of the [7,3,4] classic code. Because of this structure, it is possible to see the following property through a simple code inspection. Given the bit-flip error $X_v$, $W_H(v) = 2$, $\exists u \in GF(2)^7$ with $W_H(u) = 1$ such as $X_v |0_E\rangle = X_u |1_E\rangle$, the inverse is also satisfied. This fact will permit us not to worry about the phase errors in the $|0_E\rangle$ synthesis. Consider the network preparing the $|0_E\rangle$ state is not phase-flip fault-tolerant, meaning that more than one error could be introduced into the state. Suppose the phase error is $Z_v$ with $W_H(v) \geq 2$ and $Z_v |0_E\rangle = H X_v (H |0_E\rangle) = H X_v (|0_E\rangle + |1_E\rangle) = H X_u (|1_E\rangle + |0_E\rangle) = Z_u |0_E\rangle$. The conclusion is, there is no phase errors of weight bigger than one, so any network synthesizing $|0_E\rangle$ will be phase-flip fault-tolerant. We only have to be concerned about the bit-flips. In order to avoid their accumulation into the $|0_E\rangle$ state, an easy method to synthesize a $|0_E\rangle$ state fault-tolerantly is using a network built from the generating matrix of the [7,3,4] classic code not worrying whether it is not fault-tolerant. A second $|0_E\rangle$ state will be prepared with the same method and, finally, both states will be connected by means of a transversal CNOT gate and the second $|0_E\rangle$ state measured and collapsed onto the $|w\rangle$ sate. The quantum state achieved $|w\rangle$, considered as a classic register $w \in GF(2)^7$ will be a codeword of the [7,3,4] code, correcting bit-flips of weight one and detecting those of weight two and three. Fortunately, when this code is used to detect errors in the $|0_E\rangle$ state, can detect bit-flip errors of *any weight*, because no more than errors of weight three exist for the $|0_E\rangle$ state. When an error is detected in the measured codeword w, the whole synthesizing method is restarted. The simple synthesizing $|0_E\rangle$ network is shown in figure 10. Phase-flips appear anywhere and are back spread by means of the CNOT gates, but they always produce errors of weight one. Note that some bit-flips happen just after the Hadamard gates have no effect because they only interchange the states. If the second $|0_E\rangle$ state has no bit-flip errors, it can detect a bit-flip error of *any weight* in the first $|0_E\rangle$. The error probability per time step and gate application is $O(\varepsilon)$ and $O(\gamma)$, respectively, then the probability of this method failing to produce a correct $|0_E\rangle$ state, will came from two (or more) errors in the network, having an error probability $O(\varepsilon^2,\gamma^2)$ and being fault-tolerant.

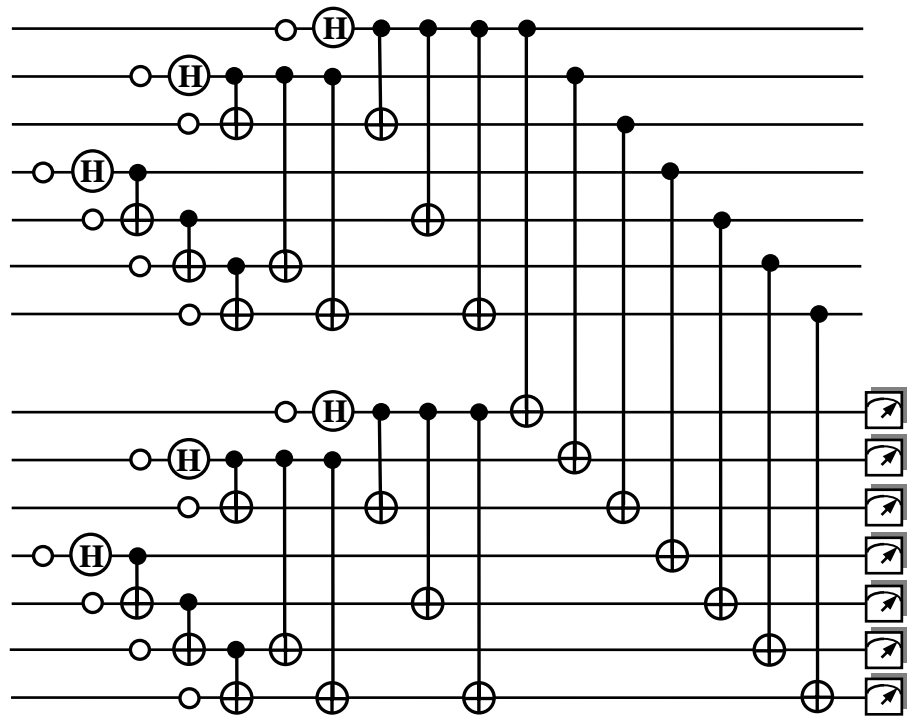

Figure 10. Detailed network for the $|0_E\rangle$ synthesis corresponding to the pieces of the network inside the dashed boxes in figure 8.

The Grover algorithm is carried out encoding the initial $|00\rangle$ state fault-tolerantly with the code [[7,1,3]] and the previous method, and the results compared with the case without encoding. The error model characteristics are kept the same as they were introduced in section 3. In the case of n=2, the result of the error free algorithm is the state $|0_E 0_E\rangle$ after one application of the Grover gate. So the definition of the error probability is the probability of obtaining a different state of the $|00\rangle$ state after the final classic correcting and decoding step. Calculations have been carried out for C = 1. and 2. and are shown in figure 11. In both cases there is an error region in which there is a passive stabilization coming from the encoding. The region in which the probability of success is greater, decreases as the error increases. Some improvement could be reached L-concatenating the code as $[[7^L,1,3^L]]$ and keeping the structure of the method. Unfortunately this will be very expensive from the experimental point of view as the number of qubits increases. To actually improve the results, an active correction or some other passive method based on a different strategy would be needed.

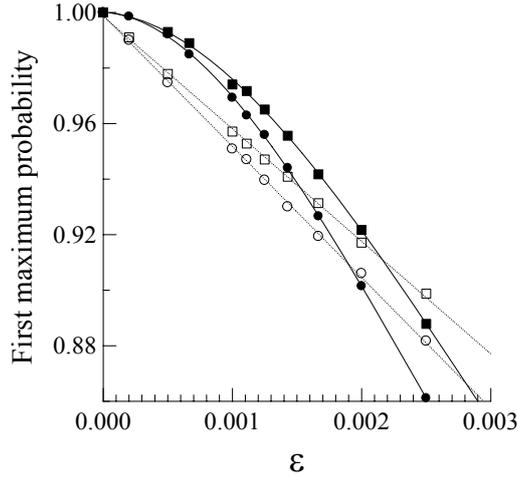

Figure 11. Comparison of the first maximum probability versus ε when encoding is used for the Grover algorithm with n=2 and different values of C=ε/γ. Dashed lines and open symbols are the results without encoding and solid lines and full symbols with encoding: (○, ●) C=1 and (□, ■) C=2.

## 5. CONCLUSIONS

Grover algorithm has been studied in presence of isotropic depolarising noise. The numerical simulation permits us conclude an exponential damping law for the successive probability of the maxima as the time increases. The iteration time at which the maximum appears depends significantly on the intensity of the noise through the error parameters. Surprisingly, the greater the noise, the fewer number of time steps are required to reach the first maximum, so less computational effort is needed to reach the item. This fact originated because the damping increases over time. Note that the corresponding probability of success for the first (and subsequent) maximum decreases. We have been able to obtain an allowed-error law for the algorithm. The error threshold for the allowed noise behaves like $\varepsilon_{th}(N) \sim 1/N^{1.1}$ as the size of the data set $N = 2^n$ (n is the number of qubits) increases. As the power of N in this law is 1.1 (near 1.) we consider the Grover algorithm as robust to a certain extent against decoherence. This law also provides an absolute threshold for the present implementation of the algorithm: if $\varepsilon > \varepsilon_{th}^{max} = 0.043$, Grover algorithm does not work for any number of qubits affected by the present error model.

Unfortunately, as the interesting problems to be solved using the Grover algorithm involve several tens of qubits, the noise thresholds to be reached by any experimental device will be very small, then some method of controlling decoherence should be necessary. We have used the binary [[7,1,3]] quantum error correcting code as a passive method to encode a two qubit state without correction. For the case of two-qubit encoded Grover algorithm, the encoding shows a region in which it is possible to increase the probability of success.

It is necessary to remark that the conclusions achieved depend on the error model considered. However, although the circuit of implementation can be optimized, we presume that the scale laws for the allowed error are still valid.

## ACKNOWLEGMENTS

The author would like to thank the Spanish Research Project CCG06-UPM/INF-389 for its financial support.

1. L. K. Grover, Phys. Rev. Lett. **78**, 325 (1997).
2. I.L. Chuang et al., Nature **393**, 143 (1998); L. M. K. Vandersypen et al., Appl. Phys. Lett. **76**, 646 (2000); G. L. Long and L. Xiao, J. Chem. Phys. **119**, 8473 (2003); M. S. Anwar et al., Chem. Phys. Lett. **400**, 94 (2004); K.-A. Brickman et al., Phys. Rev. A **72**, 050306(R) (2005); L. Xiao and J. A. Jones, Phys. Rev. A **72**, 032326 (2005).
3. B. Pablo-Norman and M. Ruiz-Altaba, Phys. Rev. A **61**, 012301 (1999).
4. S. Yu and Ch. Sun, e-print arXiv: quant-ph/9903075 (1999).
5. G. L. Long, Y. S. Li, W. L. Zhang and Ch. C. Tu, Phys. Rev. A **61**, 042305 (1999).
6. J. Hsieh, Ch. Li and D. Chuu, Chinese Jour. of Physics **42**, 585 (2004).
7. D. Shapira, S. Mozes and O. Biham, Phys. Rev. A **67**, 042301 (2003).
8. J. Chen, D. Kaszlikowski, L. C. Kwek, C. H. Oh, e-print arXiv: quant-ph/0102033 (2001).
9. P. H. Song and I. Kim, Eur. Phys. J. D **23**, 299 (2003).
10. D. Ellinas and Ch. Konstadakis, e-print arXiv:quant-ph/0110010 (2001).
11. H. Azuma, Phys. Rev. A **65**, 042311 (2002).
12. N. Shenvi, K. R. Brown and K. B. Whaley, Phys. Rev. A **68**, 052313 (2003).
13. A. A. Pomeransky, O. V. Zhirov and D. L. Shepelyansky, in *Proceedings of ERATO Conference on Quantum Information Science, 2004,* (Tokyo, 2004), pp. 171; A. A. Pomeransky, O.V. Zhirov and D.L. Shepelyansky, Eur. Phys. J. D **31**, 131 (2004).
14 O.V. Zhirov and D. L. Shepelyansky, Eur. Phys. J. D **38** 405 (2006).
15. E. Knill, R. Laflamme and W. H. Zurek, Science **279**, 342 (1998).
16. E. Knill, R. Laflamme and W. H. Zurek, Proc. R. Soc. London A **454**, 365 (1998).
17. M. Lüscher, Comp. Phys. Comm., **79**, 100 (1994).
18. F. James, Comp. Phys. Comm., **79**, 111 (1994).
19. P.J. Salas and A.L. Sanz, Int. J. Quantum. Inf., **3**, 371 (2005).